\documentclass[a4paper,aps,pra,preprintnumbers,amsmath,amssymb,twocolumn,longbibliography,superscriptaddress]{revtex4-1} 
 \pdfoutput=1 
 
\usepackage{physics}
\usepackage{bbold}
\usepackage{mathtools}
\usepackage[utf8]{inputenc}
\usepackage[T1]{fontenc}
\usepackage{graphicx}
\graphicspath{{figures/}}
\usepackage{amssymb}
\usepackage{amsmath}
\usepackage{algorithmic}
\usepackage{array}
\usepackage{hyperref}

\usepackage{xcolor} 

\begin{document}

\title{Preparation of the 1/2-Laughlin state with atoms in a rotating trap}

\author{Bárbara Andrade}
\affiliation{Instituto de Física Teórica, UNESP-Universidade Estadual Paulista, São Paulo 01140-070, SP, Brazil}
\affiliation{ICFO-Institut de Ciencies Fotoniques, The Barcelona Institute of Science and Technology, 08860 Castelldefels (Barcelona), Spain}

\author{Valentin Kasper}
\affiliation{ICFO-Institut de Ciencies Fotoniques, The Barcelona Institute of Science and Technology, 08860 Castelldefels (Barcelona), Spain}
\affiliation{Department of Physics, Harvard University, Cambridge, MA, 02138, US}

\author{Maciej Lewenstein}
\affiliation{ICFO-Institut de Ciencies Fotoniques, The Barcelona Institute of Science and Technology, 08860 Castelldefels (Barcelona), Spain}
\affiliation{ICREA, Pg. Lluis Companys 23, 08010 Barcelona, Spain}

\author{Christof Weitenberg}
\affiliation{ILP — Institut für Laserphysik, Universität Hamburg, Luruper Chaussee 149, 22761 Hamburg, Germany}
\affiliation{The Hamburg Centre for Ultrafast Imaging, Luruper Chaussee 149, 22761 Hamburg, Germany}

\author{Tobias Gra\ss}
\affiliation{ICFO-Institut de Ciencies Fotoniques, The Barcelona Institute of Science and Technology, 08860 Castelldefels (Barcelona), Spain}

\begin{abstract}

Fractional quantum Hall systems are among the most exciting strongly correlated systems. Accessing them microscopically via quantum simulations with ultracold atoms would be an important achievement toward a better understanding of this strongly correlated state of matter. A promising approach is to confine a small number of bosonic atoms in a quasi-two-dimensional rotating trap, which mimics the magnetic field. For rotation frequencies close to the in-plane trapping frequency, the ground state is predicted to be a bosonic analog of the Laughlin state. Here, we study the problem of the adiabatic preparation of the Laughlin state by ramping the rotation frequency and controlling the ellipticity of the trapping potential. By employing adapted ramping speeds for rotation frequency and ellipticity, and large trap deformations, we improve the preparation time for high-fidelity Laughlin states by a factor of ten
in comparison to previous studies. With this improvement of the adiabatic protocol the Laughlin state can be prepared with current experimental technology.

\end{abstract}

\maketitle

\section{Introduction}

 Ultracold atoms give a unique perspective on strongly correlated matter \cite{Bloch2005,lewenstein2012ultracold} as they allow one, for example, to study quantum states with single-atom resolution or to explore higher-order correlations and entanglement \cite{Kaufman2016, Schweigler2017}. Moreover, ultracold atoms have several features, which make them particularly well suited for the study of strongly correlated matter. Their isolation from the environment is excellent and the microscopic system parameters are highly tunable.  This tunability allows for preparing a variety of strongly correlated states by adiabatically ramping the system parameters starting from a well-defined state such as a trapped Bose-Einstein condensate.

Strongly correlated states of particular interest are fractional quantum Hall states, especially because of their prospects for topological quantum computation \cite{Kitaev2003}. Although fractional quantum Hall physics has been experimentally discovered already four decades ago \cite{tsui82}, and has readily been explained in terms of Laughlin's trial wave function \cite{laughlin83}, the fractional quantum Hall effect continues to be a challenging subject of research: One of the most striking predictions about the fractional quantum Hall physics is the existence of quasiparticles with fractional statistics \cite{leinaas77,wilczek82}, so-called anyons. The existence of these quasi-particles has yet to be confirmed ultimately, despite strong efforts and much experimental progress made towards anyon detection \cite{saminadayar97,camino05,camino07,bartolomei20}. 

A new direction of how to approach these challenges are quantum simulators, which prepare fractional quantum Hall states in highly controlled experimental settings. Many advances towards such synthetic fractional quantum Hall systems have been made in both atomic \cite{miyake13,aidelsburger13,aidelsburger15,flaeschner16,asteria19,tarnowski19} and photonic \cite{hafezi13,rechtsman13,mittal14,mittal16,bandres16,baboux17} quantum simulators. These advances include the generation of artificial magnetic fields, which are responsible for the flat band structure, and detection of their topological properties, such as chiral edge states \cite{hafezi13,rechtsman13}, topological quantum numbers \cite{aidelsburger15,mittal16,baboux17,asteria19,tarnowski19}, topological transport \cite{mittal14,bandres16}. Through light-matter coupling, it has also been possible to create interactions between two photons in a synthetic gauge field, yielding a Laughlin-type quantum state \cite{clark20}. Although atomic systems are interacting in a more natural way, the evidence of atomic Laughlin states has remained limited until now \cite{gemelke10}.

Various difficulties in reaching synthetic Laughlin states are known: In the strongly correlated regime, the centrifugal forces leading to the artificial gauge field almost compensate the trap \cite{dagnino09,julia-diaz11}, and thus reduces the stability of the system. Adding steeper potentials to the harmonic trap such as a confining quartic potential or a weak hard wall confinement have been found to be very harmful to bosonic Laughlin states \cite{roussou19, Macaluso2017}. 
The generation of synthetic gauge fields may heat the system, especially if periodic driving is involved \cite{dalessio14}. In this context, it is particularly important to note that various intermediate phases separate the uncorrelated system from the strongly correlated liquid phase \cite{viefers00,viefers08,dagnino09,julia-diaz11}. Thus, the phase diagram exhibits different regions of small energy gaps above the ground state. Nevertheless, an adiabatic path to the Laughlin state has been proposed for a system of bosonic cold atoms in a harmonic elliptic trap with tunable rotation frequency and tunable ellipticity \cite{popp04}. Similar considerations for the adiabatic preparation also apply to fermionic systems \cite{Palm2020}. The adiabatic preparation scheme can also be applied to systems in rotating ring potentials \cite{Roncaglia2011}. Another route to synthetic Laughlin states is based on ``growing'' the state via variable particle numbers \cite{grusdt14}.

In the present paper, we revisit the adiabatic preparation scheme for bosonic Laughlin states in rotating traps \cite{popp04}. The idea is to increase the angular momentum $L$ of $N$ atoms in a rotating trap from the non-rotating state $L=0$ to the angular momentum of the 1/2-Laughlin state, $L=N(N-1)$, by a ramp of the rotation frequency of the trap, and simultaneously breaking rotational symmetry by an anisotropic deformation of the trap. In Ref.~\cite{popp04}, a preparation time of $6450$ trapping periods was reported, in which the Laughlin state of four atoms was reached with a fidelity of 0.97. This implies that even for a trapping frequency as large as $(2\pi)\times30$~kHz, the preparation time exceeds 200~ms. However, we show that such an adiabatic preparation can dramatically be improved. Specifically, our numerics reach a four-atom Laughlin state with a fidelity of 0.99 within $605$ trapping periods, or 20~ms for a frequency of $(2\pi)\times30$~kHz. This result significantly improves the prospects of preparing atomic Laughlin states using an adiabatic scheme. The main ingredients that distinguish our scheme from earlier work are:
\begin{itemize}
\item  larger anisotropies of the trap: During the preparation the atoms acquire large values of angular momentum, exceeding the Laughlin value, far before reaching the strongly correlated regime. Thus, the accumulation of angular momentum occurs in regimes which are characterized by relatively large energy gaps, and in the final stage of the protocol, the Laughlin state is approached by \textit{reducing} the angular momentum of the system.
\item varying ramp speeds: relatively large energy gaps allow for quick ramps at an early stage of the preparation scheme, shortening the total evolution time.
\end{itemize}
Our work is organized as follows: In Sec.~\ref{System}, we describe the system and its behavior at different rotation frequencies. In Sec.~\ref{Results} we present how rotation frequency and trap anisotropy can be tuned to reach the Laughlin state with high fidelity. In Sec.~\ref{sec:robustness} we comment on the robustness of the proposed protocol. Conclusions of this result are drawn in Sec.~\ref{Conclusions}.

\section{Theoretical Model} \label{System}
We consider a microscopic model of $N$ bosonic atoms confined to two dimensions and trapped in a harmonic potential. These microtraps can be realized either via a tightly-focused optical tweezer or via an optical lattice as a decoupled array of individual microtraps as in Ref.~\cite{gemelke10}. Tight harmonic confinement along the third dimension ($z$-direction) freezes all excitations along that direction, and each microtrap becomes effectively two-dimensional. We denote the harmonic oscillator frequency by  $\omega_z$, and the associated length scale is given by $\lambda_z = (\hbar/M\omega_z)^{1/2}$, with $M$ the mass of the atoms. The bosonic atoms interact via contact interaction, which we parametrize
with the dimensionless coupling constant $g$. In the considered experimental setups
the dimensionless coupling is given by $g = \sqrt{8\pi} (a_S/\lambda_z)$, with $a_S$ being the three-dimensional scattering length.
The artificial gauge field is created by rotation around the $z$ direction with frequency $\Omega$. For a review on artificial gauge fields with atoms in a rotating trap, we suggest Refs. \cite{Cooper2008, Fetter2009}. The total Hamiltonian $H = H_0 + H_I$ describing $N$ atoms consists of the non-interacting part
\begin{align}
    H_0 = \sum_{j=1}^N \left[ \frac{\mathbf{p}_{j}^{\,2}}{2M} +\frac{1}{2}M\omega^2 \mathbf{r}_{ j}^{\:2}-\Omega L_{z,j} \right] \, ,
\end{align}
and the interacting part
\begin{align}
    H_I =  \frac{\hbar^2g}{M}\sum_{j=1}^N\sum_{k>j}\delta(\mathbf{r}_{j}-\mathbf{r}_{k}) \, ,
\end{align}
where $\mathbf{r}_{j} = x_j \mathbf{e}_x + y_j \mathbf{e}_y$ is the position operator in the $xy$-plane, and $L_{z,j}$ is the angular momentum operator in $z$-direction of the $j$th atom. Moreover, $\omega$ is the frequency of the harmonic trapping in the $xy$-plane. The single particle
Hamiltonian can be written as
\begin{equation}
H_0 = \sum_{j=1}^N   \left[ \frac{|\mathbf{p}_{j}-M \mathbf{\Omega} \times \mathbf{r}_{ j}|^{2}}{2 M}+\frac{1}{2} M\left(\omega^{2}-\boldsymbol{\Omega}^{2}\right)\mathbf{r}_{ j}^{\:2}  \right]\!,
\label{eq:ParticleInGaugeField}
\end{equation}
where we introduced the rotation vector $\boldsymbol{\Omega}=\Omega \hat{z}$ along the z-axis.
Eq.~\eqref{eq:ParticleInGaugeField} describes non-interacting particles with charge $q$ 
in a magnetic field $q \mathbf{B} = 2 M \boldsymbol{\Omega}$.

The single-particle eigenstates of $H_0$ are the Fock-Darwin states, cf. Ref.~\cite{dagnino09}, which are organized in different Landau levels, separated by a ``cyclotron'' energy $\hbar(\omega+\Omega)$. Different states within a Landau level are distinguished by an angular momentum quantum number $m$, which contributes the term $m\hbar(\omega-\Omega)$ to the single-particle energy. Assuming that $\omega+\Omega \gg \omega-\Omega$, and that the cyclotron energy also sufficiently exceeds the interaction energy of the system, the effective Hilbert space can be reduced to the lowest Landau level. The Fock-Darwin wave functions of the lowest Landau level are given by
\begin{align}
    \phi_m(x,y)=\frac{1}{\lambda^{m+1}\sqrt{\pi m!}}\:(x+iy)^me^{-(x^2+y^2)/2\lambda^2},
\end{align}
where $\lambda=\sqrt{\frac{\hbar}{M\omega}}$ is the harmonic oscillator length scale.

 We use these eigenstates as a computational basis. The second-quantized operator $a^\dagger_m$ ($a_m$) creates (annihilates) a particle described by  $\phi_m(x,y)$. Expressing energies in units of $\hbar\omega$, frequencies in units of $\omega$, and angular momentum in units of $\hbar$,  in second quantization the Hamiltonian can be written as
\begin{align}
    H = H_0 + H_I = {N} + [1-\Omega]{L} + {U},
\end{align}
where ${N}=\sum_{m}a^\dagger_ma_m$ is the number operator, ${L}=  \sum_{m}  m a^\dagger_ma_m$ is the total angular momentum operator (in units of $\hbar$), and ${U}= H_I/(\hbar \omega)$ is the interaction operator
\begin{align}
    {U}&=\sum_{m,n,p,q}U_{m,n,p,q}\:a^\dagger_ma^\dagger_na_pa_q \,,
\end{align}
where the matrix element is given by
\begin{align}
    U_{m,n,p,q}&=\frac{g}{\pi}\frac{\delta_{m+n,p+q}}{\sqrt{m!n!p!q!}}\frac{(m+n)!}{2^{m+n+1}}.
\end{align}
All terms in the Hamiltonian commute with ${L}$, and hence the angular momentum is a conserved quantity at this point.

We are interested in preparing the ground state of a bosonic fractional quantum Hall system at Landau filling fraction $\nu=1/2$, i.e. the lowest Landau level shall be half-filled. For particles which interact with short-range interactions such phase is exactly described by the $1/2$-Laughlin wavefunction
\begin{align}
    \psi_{L}(z_1,\ldots, z_N)=\prod_{i<j}^{N}\left(z_{i}-z_{j}\right)^{2} \prod_{k=1}^{N} e^{-\left|z_{k}\right|^{2} / 2}.
    \label{laughlin}
\end{align}
Here, we have used complex numbers $z_j$ to represent the position of the $j$th particle, $z_j = (x_j + iy_j)/\lambda$. This symmetric wave function is zero whenever two particles are at the same position, and thus, it is a zero-energy eigenstate of the contact potential $H_I$.

The 1/2-Laughlin state has total angular momentum $L=N(N-1)$ (in units $\hbar$), as can be inferred from the degree of the polynomial part of Eq.~\eqref{laughlin}. On the other hand, the total angular momentum of the ground state of $H$ is the result of a competition between $H_0$ and $H_I$: The single-particle part $H_0$ yields an energy which is proportional to $L$, while larger values $L$ allow the particles to avoid each other, reducing the amount of interaction energy. In particular, there are no zero-energy eigenstates of $H_I$ for $L<N(N-1)$.
We can control this competition of $H_0$ and $H_I$ by the rotation frequency in $H_0$, which in the following will therefore be chosen to be time-dependent, i.e. $\Omega(t)$. Throughout the paper, we will express $\Omega(t)$ in units of $\omega$.

\begin{figure}[t] 
  \includegraphics[scale=1.15]{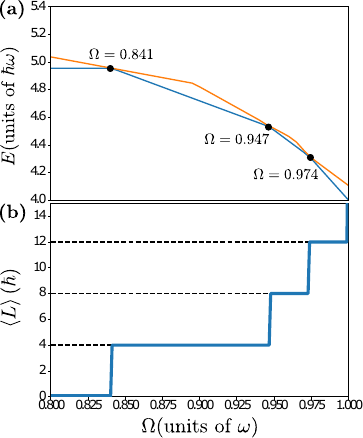} 
  \caption{\textbf{(a)} Energy of ground state and first excited state in an isotropic system of four atoms (with $g=1$) as a function of rotation frequency. True level crossings happen at $\Omega=0.841,0.947$ and $0.974$. \textbf{(b)} Average angular momentum of the ground state as a function of rotation frequency. The Laughlin state is the ground state after the third crossing, when $L=N(N-1)=12$.} 
  \label{A=0}
\end{figure}

This competition is illustrated in Fig.~\ref{A=0}, where we have plotted the energy of ground state and first excited state in Fig.~\ref{A=0}a, and the total angular momentum of the ground state in  Fig.~\ref{A=0}b as a function of the rotation frequency $\Omega$. At discrete values of $\Omega$, the energy gap above the ground state vanishes, and the ground state angular momentum changes abruptly. In the system of four particles, we obtain ground states at $\expval*{{L}}=0,4,8$ and $12$. It will be the goal of our adiabatic protocol to bring a rotating system from the condensate phase ($L=0$) to the Laughlin state ($L=N(N-1)$) by a ramp of the rotation frequency. In this work, we consider the experimentally relevant case of $N=4$ atoms implying an angular momentum of $L=12$ for the Laughlin state. We fix the interaction parameter to $g=1$, noting that in practice $g$ can be tuned via Feshbach resonances and/or confinement-induced resonances. 

The transitions in Fig.~\ref{A=0} are true level crossings, as allowed by the rotational symmetry of the system. 
Thus, in order to adiabatically connect the different ground states, we have to turn these true crossings into avoided crossings. This can be achieved by removing the rotational symmetry, e.g. by introducing an anisotropic potential to the Hamiltonian
\begin{align}
  {V}(t) = A(t) M\omega^2 \sum_i ({x}_i^2-{y}_i^2) 
\end{align}
or, in terms of creation and annihilation operators and in untis of $\hbar \omega$,
\begin{align}
    {V}(t) = \frac{A(t)}{2}\sum_{m=2}^{\infty} \left[ \sqrt{m(m-1)}{a}_m^\dagger {a}_{m-2}+ {\rm h.c.} \right].
\end{align}
With this, the new Hamiltonian for the system is
\begin{align}
    H(t) = {N} + [1-\Omega(t)] {L} +  {U} +  {V}(t).
\end{align}

These expressions for $V(t)$ implicitly define an ``anisotropy'' parameter $A(t)$, which together with the rotation frequency $\Omega(t)$  shall be controllable as a function of time. Our goal is to fix the temporal behavior of these parameters such that the system evolves into the Laughlin state. We note that the anisotropy in $V(t)$ is due to an increase of the trapping frequency along the $x$-direction, and a decrease of the trapping frequency along the $y$-direction. Concretely, the trapping frequency along $y$-direction is proportional to $\sqrt{1-2A}$, which sets the centrifugal limit to $\Omega\leq\sqrt{1-2A}$. For larger rotation frequencies, the state preparation is expected to become more delicate since atoms can be expelled from the trap. We will avoid this region in our protocol.

\begin{figure}[t] 
  \includegraphics[scale=0.90]{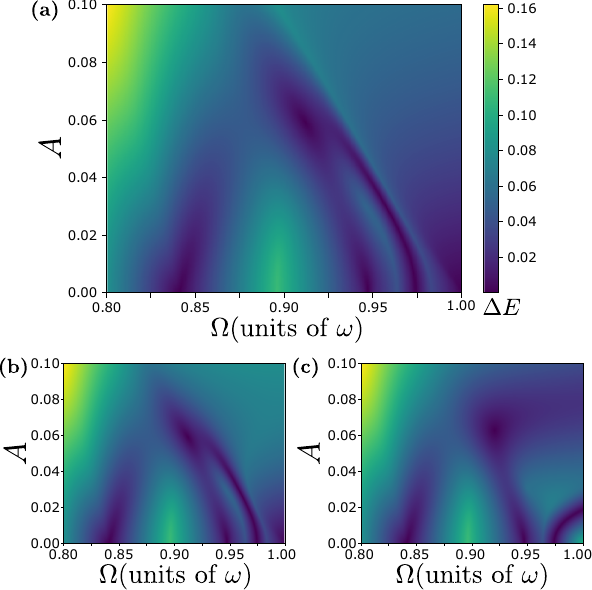}
  \caption{\textbf{Energy gap} as a function of rotation frequency and anisotropy parameter for different angular momentum truncations: \textbf{(a)} $L_{\rm max}=40$ \textbf{(b)} $L_{\rm max}=26$  \textbf{(c)} $L_{\rm max}=12$  All plots share the same color scale as (a), the energy gap $\Delta E$ is given in units of $\hbar\omega$.}
  \label{colorplots}
\end{figure}

The anisotropy also introduces additional complexity from the computational point of view: Since the new Hamiltonian does not conserve the total angular momentum, we must truncate the Hilbert space at some $L=L_{\rm max}$. The choice of $L_{\rm max}$ depends on the protocol. More precisely, in order to have good convergence of our simulations we must assure that, at all times, the sectors of large $L$ (i.e. close, equal and above $L_{\rm max}$) contribute a negligible part to the many-body wavefunction. In Fig.~\ref{colorplots}, we plot the energy gap above the ground state as a function of anisotropy parameter $A$ and rotation frequency $\Omega$ for different choices of $L_{\rm max}$. This comparison illustrates that truncation at fairly small values, such as $L_{\rm max}=12$ in Fig.~\ref{colorplots}(c), is possible only for small values of $A$ or $\Omega$. On the other hand, in Fig.~\ref{colorplots}(a) and (b), the energy gap for $L_{\rm max}=26$ and $L_{\rm max}=40$ agree very well in the whole parameter region, suggesting that good convergence of the numerics has been reached. For our simulation of the adiabatic state preparation, presented in the next section, we have chosen $L_{\rm max}=40$. This truncation provides good convergence in the protocol we propose for the Laughlin state preparation.

\section{Adiabatic State preparation  \label{Results}}

In this section, we study a specific protocol for $A(t)$ and $\Omega(t)$ which adiabatically moves the system from the condensate ($L=0$) into the Laughlin state ($L=12$). In order to ensure adiabaticity, regions with small energy gap should be avoided, while the velocity of parameter changes should be adjusted to the size of the energy gap. At the same time, in order to facilitate the implementation of the protocol, we want to keep the parameter speed constant along extended pieces of the path.

\begin{figure*}
  \includegraphics[scale=1]{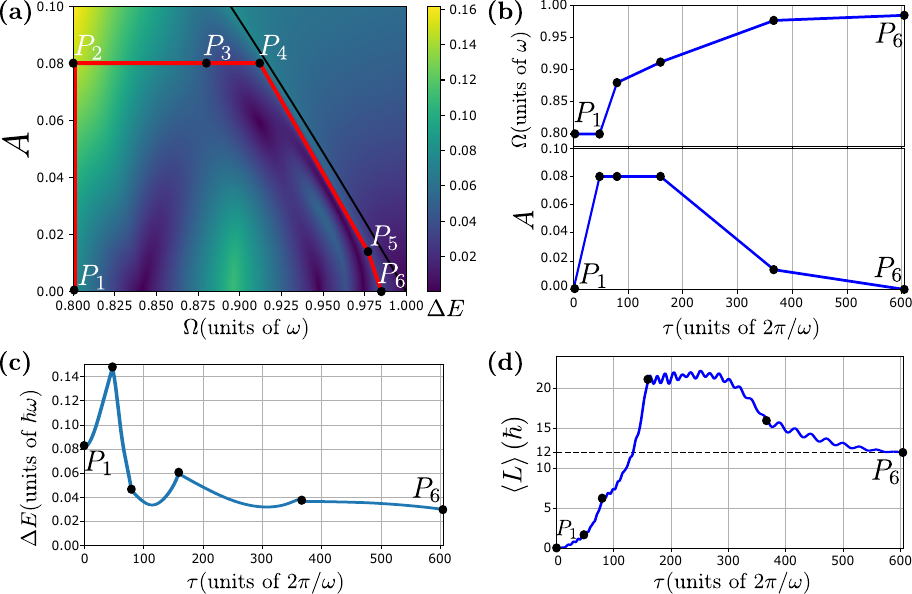}
  \caption{Characteristic of adiabatic Laughlin state preparation. \textbf{(a)} Path in the parameters space for truncation $L_{\rm max}=40$. The black line is defined by $\Omega=\sqrt{1-2A}$, which bounds the region where the preparation becomes more delicate. \textbf{(b)} Energy gap along the protocol. \textbf{(c)} Rotation frequency and anisotropy parameter as a function of time. \textbf{(d)} Average angular momentum as a function of time. The precise coordinates of the points and time marks are given in Table \ref{points}, in (b), (c) and (d) we omit the label of intermediate points for better visualization.}
  \label{protocol}
\end{figure*}

With these considerations in mind, we have considered the protocol as illustrated by the red line in Fig.~\ref{protocol}(a): First, the anisotropy is ramped up to a relatively large value ($A=0.08$) at slow rotation ($\Omega=0.8$). Next, the rotation frequency is increased almost up to the centrifugal limit (marked by the black line in Fig.~\ref{protocol}). Finally, we simultaneously decrease $A$ and increase $\Omega$ along the centrifugal limit, until isotropy is restored and the Laughlin state is reached. From the contour plot of the energy gap, it is obvious that this path avoids regions of small gaps.

Furthermore, we allocate different amounts of time for the evolution along different segments of the path. To this end, we have marked different points $P_i=(\Omega_i,A_i)$ along the path, which shall be reached at given times $t_i$. Between adjacent points, the parameters $A(t)$ and $\Omega(t)$ are changed linearly in time. Thus, the protocol is fully determined by $P_i$ and $t_i$, as given by Table~\ref{points}. In this table, we have parametrized time $t$ by dimensionless values $\tau=\frac{\omega\:t}{2\pi}$, which measure time in units of the trapping period. An illustration of the protocol defined by Table~\ref{points} is provided in Fig.~\ref{protocol}(b). With the chosen timing, our protocol is significantly slowed down in the regions of small gap (between $P_3$ and $P_4$, and between $P_5$ and $P_6$), while it quickly passes the other regions. This can also be seen from Fig.~\ref{protocol}(c), which plots the energy gap as a function of $\tau$.

\begin{table}[h!]
\centering
\begin{tabular}{| c | c | c | c |c|} 
    \hline
    & $\Omega_i$ & $A_i$ & $\tau_i$ & $\Delta \tau_i$ \\ 
    \hline
    $P_1$ & $0.8$   & $0$    & $0$ & - \\ 
    $P_2$ & $0.8$   & $0.08$  & $48$ & 48\\
    $P_3$ & $0.88$  & $0.08$  & $80$ & 32\\
    $P_4$ & $0.912$ & $0.08$  & $160$ & 80\\
    $P_5$ & $0.977$ & $0.014$ & $366$ & 206\\
    $P_6$ & $0.985$ & $0$     & $605$ & 239\\
    \hline
\end{tabular}
\caption{Coordinates $(\Omega_i,A_i)$ of the points $P_i$ along the protocol in Fig.~\ref{protocol}(a), and the dimensionless time $\tau_i$ at which the given configuration is reached within the protocol. The difference $\Delta \tau_i=\tau_i-\tau_{i-1}$ measures the amount of time spent to evolve between adjacent points.}
\label{points}
\end{table}

A measure for the adiabatic nature of the evolution is the fidelity $F(\tau)$ as a function of time, defined as the squared overlap between the evolved state at time $\tau$ with the instantaneous ground state of the Hamiltonian $H(\tau)$. At the end of the protocol, this quantity becomes  the fidelity with which the Laughlin state is reached, i.e. a measure for the quality of the protocol. Fixing the total evolution time at $T=605$ (in units $2\pi/\omega$), our protocol reaches the Laughlin state with fidelity $F(T)=0.99$, and during the evolution, the ``instantaneous'' fidelity $F(\tau)$ always remains above $F>0.98$. These numbers indicate that the protocol operates with good approximation in an adiabatic regime.

The chosen evolution time, $T=605$, corresponds to 20~ms, 60~ms and 200~ms for trapping frequencies of $\omega$ = $(2\pi)\times$ 30 kHz, 10 kHz and 3 kHz, respectively. The total time 
for the Laughlin state preparation appears to be in an experimental accessible regime. However, the frequencies only correspond to the in-plane trap, whereas the trapping frequency along $z$ must be chosen much larger than $\omega$, which sets experimental limitations.

Naturally, the angular momentum reached at the end of the protocol is very close to the desired value, $L=12.02$. However, it is noteworthy that this value is not reached by a monotonous increase of $L$. In Fig.~\ref{protocol}(d), we see that significantly larger values of $\langle L \rangle >20$ are reached when the system is closest to the centrifugal limit, i.e. between $P_4$ and $P_5$. Only in the very end, between $P_5$ and $P_6$, our protocol converges to the correct value of $12$. Therefore, although in $P_5$ the rotation frequency $\Omega$ has already the correct value for the Laughlin state, $\Omega>0.974$ as in Fig.~\ref{A=0}, one still has to decrease the ellipticity to obtain the correct angular momentum $\expval*{L}=12$.

Despite the high angular momentum values reached in the here presented protocol the Hilbert space sectors with $L>34$ are barely populated: For the instantaneous ground state along the red line in Fig.~\ref{protocol}(a), the weights of the many-body wave function \footnote{In the total angular momentum basis, a many-body state is $\ket{\Psi}=\sum_{l=0}^{L_{{\rm}} max} \sum_{j=1}^{d_l} \psi_{l,j}\ket{l,j}$, where $d_l$ is the number of states in the basis with total angular momentum $l$. For $\ket{\Psi}$ normalized, the weights $c_l=\sum_{j=1}^{d_l}\psi_{j,l}$ satisfy $\sum_{l=0}^{L_{\rm max}}c_l^2=1$.} in the $L=36$, $38$, and $40$ sectors are at most $c_{36}^2=0.012$, $c_{38}^2=0.005$, and $c_{40}^2=0.001$. The small values of the weights of the instantaneous ground states together with assumption of quasi-adiabatic preparation assure convergence of numerical simulations along this path for truncation at $L_{\rm max}=40$. Angular momentum truncation could be made at smaller values if our path was restricted to a lower anisotropy region. However, it is obvious from the contour plot of the energy gap, see Fig.~\ref{colorplots}, that smaller anisotropy values would also decrease the size of the smallest gaps along the path. Therefore, the protocol would lose fidelity very fast if we wanted to keep the same total time of $T=605$ trapping periods. A systematic study of the chosen maximum anisotropy value will be presented in the next section, in which we analyze the robustness of our results. In Appendix~\ref{AppendixPopp2004}, we compare our results to the previous study of Ref.~\cite{popp04}.

 
\section{Robustness of the protocol \label{sec:robustness}}
The previous section has considered a particular protocol $(\Omega(t), A(t))$ for fixed system parameters, demonstrating that a fast preparation of the Laughlin state is possible.  The present section studies the robustness of that protocol against variations of either the protocol itself or of the system parameters. Specifically, we investigate how using different values for interaction parameter $g$ can change the final state fidelity, the intermediate fidelity, and the average angular momentum. Moreover, we simulate protocols with different duration, or different allocation of the the time while fixing the total duration, as well as protocols with different paths (reducing the maximum trap deformation $A$).

\begin{figure}[h!] 
  \includegraphics[scale=0.55]{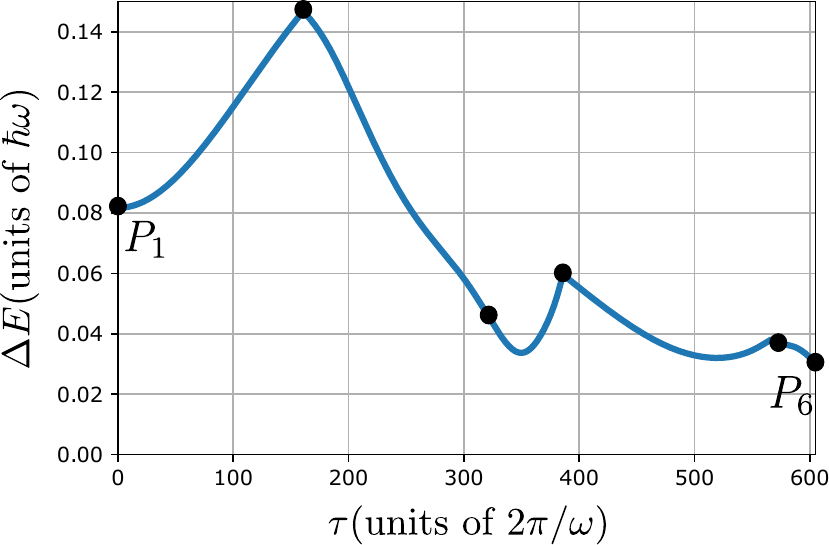}
  \caption{Gap along the red line in Fig.~\ref{protocol}(a) with homogeneous time distribution. In this case the time spent to go from $P_i$ to $P_{i+1}$ is a fraction of the total time $T=605$ proportional to the geometric distance between these points. The total time is still $T$ and the parameters are changed linearly in time on each segment.}
  \label{gapequal}
\end{figure}

\paragraph{ Protocol with constant ramp speed} In order to quantify the role of adjusted ramp speeds, we present here an alternative protocol with constant ramp speed for comparison. The path through parameter space is the same as before, but the timing is chosen as illustrated in Fig.~\ref{gapequal}. This choice is such that the time between two points, $\Delta \tau_i = \tau_i-\tau_{i-1}$, is  proportional to the geometric distance between the points $\tau_i \propto [(A_i-A_{i-1})^2+ (\Omega_{i}-\Omega_{i-1})^2]^{1/2}$, thus the protocol corresponds to homogeneous ramp speeds. With this choice, more than half of the preparation time is spent for the evolution through relatively strongly gapped regions, i.e. from $P_1$ to $P_3$, whereas in our protocol with adjusted ramp speeds defined in Table~\ref{points} the $P_1$ to $P_3$ evolution takes less than 15\% of the total protocol duration.

Whereas the protocol with adjusted ramp speed had reached the Laughlin state with fidelity $F(T)=0.99$, the new protocol with homogeneous ramp speed achieves a fidelity of $F(T)=0.94$. During the evolution, the ``instantaneous'' fidelity $F(\tau)$ now drops to values below $F<0.92$, indicating that non-negligible excitations are produced which before had been avoided by adjusting the ramp speed.

\paragraph{Shorter preparation times}

In Table~\ref{DifferentTimes}, we show how the final state fidelity responds to a decrease in the total duration of the protocol. We have simulated the same path in parameters space, but with time spent at each segment multiplied by a factor $\alpha$ between $0.1$ and $1$. Notably, the final state fidelity remains on a similar high levels if the duration is shortened by up to a factor  $\alpha=0.8$, and curiously even takes a slightly higher value than for $\alpha=1$. The fidelity drops when we decrease the total time by half, but even in this case, it still remains above $0.9$. 

\begin{table}[h!]
\centering
\begin{tabular}{| c | c |} 
    \hline
    $\rm{Duration} [T]$ & $\rm{Fidelity}$ \\ 
    \hline
    $0.1$  & $0.532$ \\
    $0.5$  & $0.901$ \\ 
    $0.8$  & $0.99$  \\
    $0.9$  & $0.983$ \\
    $0.95$ & $0.968$ \\ 
    $1$    & $0.985$ \\
    \hline
\end{tabular}
\caption{Final state fidelity for different protocol duration. All protocols are with respect to the same path in parameters space, and the duration is given in units of the original total time of $T=605$ trapping periods.}
\label{DifferentTimes}
\end{table}


\paragraph{Paths with less deformation}

\begin{figure}[t]
  \includegraphics[scale=0.4]{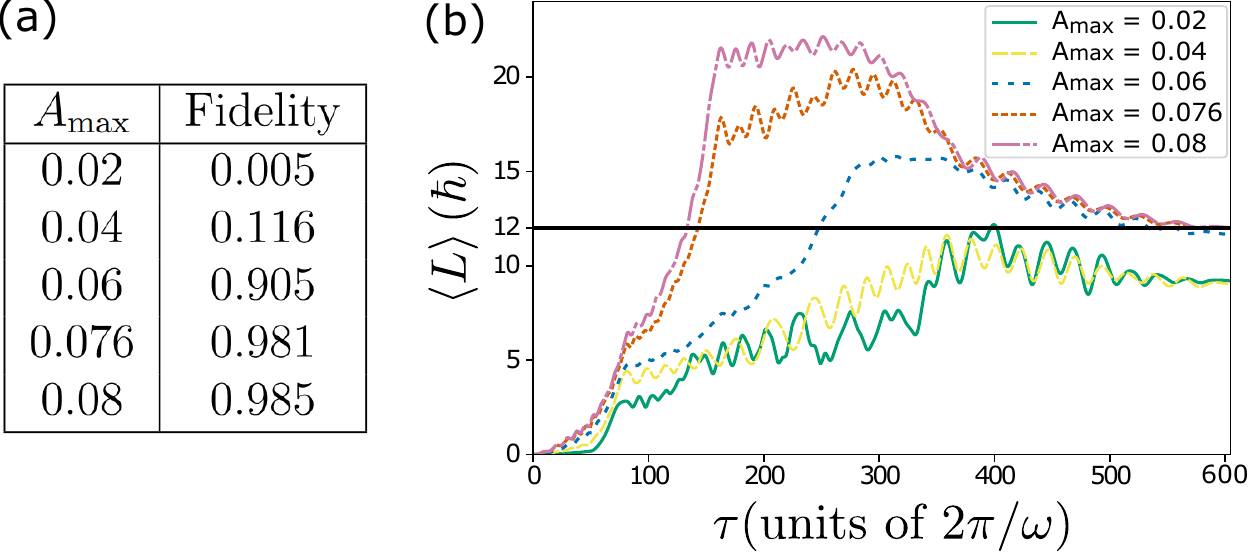}
  \caption{Results obtained from simulations of paths with less deformation. (a) Maximum trap deformation ($A_{\rm{max}}$) and final state fidelity, (b) Average angular momentum as a function of time for different values of $A_{\rm{max}}$.}
  \label{DifferentAmax}
\end{figure}

In our main result, Fig.~\ref{protocol}, the maximum trap deformation achieved was $A_{max}=0.08$. In Fig.~\ref{DifferentAmax}, we present the effect of decreasing this maximum trap deformation. Among the values we have chosen, the final state fidelity drops below $F=0.9$ only for $A_{max}=0.04$ and $A_{max}=0.02$, these protocols that do not achieve $\expval*{L}>12$ at the intermediate times. There is a strong decrease in final state fidelity because in these cases the paths in parameters space cross a region of narrow energy gap.

\paragraph{ Varying the interaction parameter}

\begin{figure}[t]
  \includegraphics[scale=0.4]{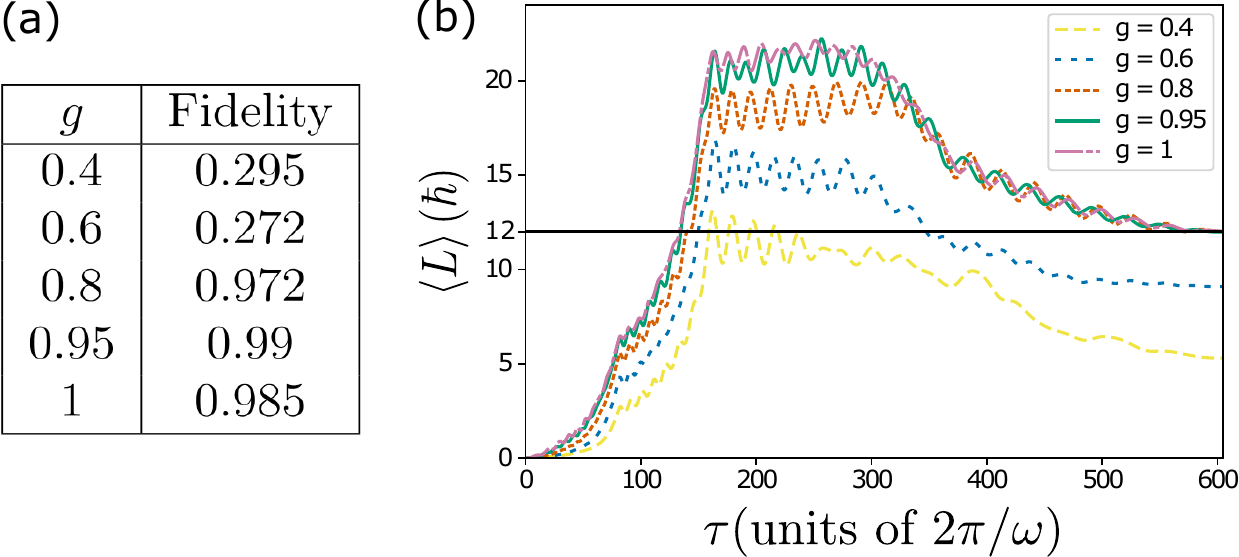}
  \caption{Results obtained from simulating paths with different interaction parameter $g$. (a) Interaction parameter and final state fidelity, (b) Average angular momentum as a function of time for different values of the interaction parameter.}
  \label{Differentg}
\end{figure}

We have also analysed the robustness of the protocol towards different interaction strengths $g$. After simulating the protocol of Table~\ref{points} for several values of $g$, we obtained the final state fidelities given in Fig.~\ref{Differentg}(a). Although the protocol has not been adjusted to the modified energy gap landscape one still observes high final state fidelity even when $g$ is 20\% weaker than the value $g=1$ used in the previous section. This calculation demonstrates that the preparation scheme still works even if system parameters are slightly miscalibrated.

\begin{figure}[t]
  \includegraphics[scale=0.2]{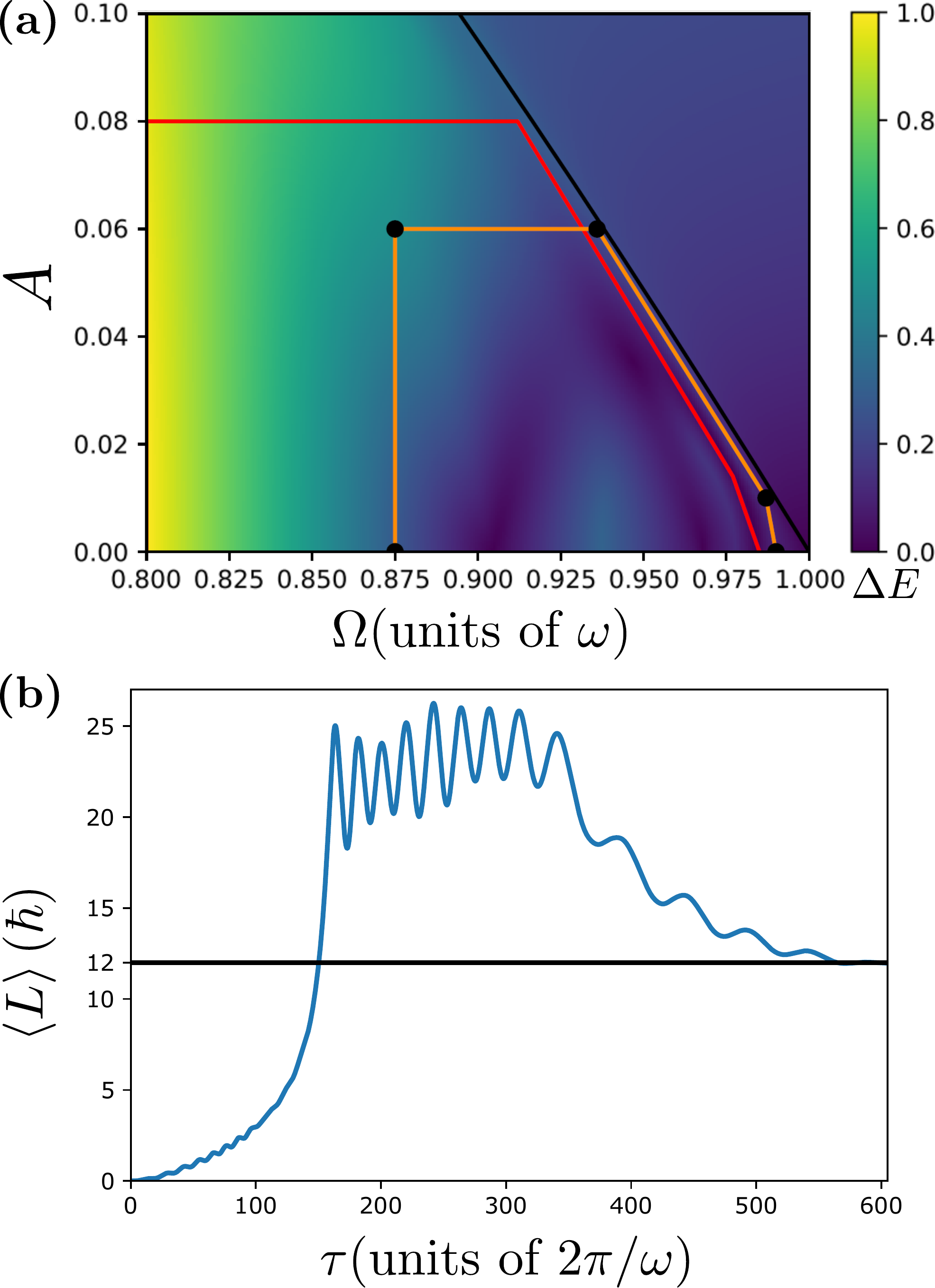}
  \caption{(a) Energy gap profile for $g=0.6$ and angular momentum truncation at $L_{\rm{max}}=40$. The orange curve is the path simulated for the Laughlin state preparation with $g=0.6$, the total duration of this protocol is $T=605$ trapping periods. The black curve, defined by $\Omega=\sqrt{1-2A}$, bounds the region where the preparation becomes more delicate. For comparison, we indicate in red the path used when $g=1$ in Fig.~\ref{protocol}(a). (b) Average angular momentum as a function of time along the protocol.}
  \label{g0.6}
\end{figure}

However, a steep drop of fidelity to $F<0.3$ occurs for $g=0.6$ or $g=0.4$, i.e., for a mismatch of $40\%$ or more from the original value. The abrupt drop in fidelity is explained by the fact that the energy gap profile changes substantially.
Indeed, this is shown in Fig.~\ref{g0.6} where we plot the energy gap for $g=0.6$.  The original path in parameters space crosses regions with  narrow gap region, but by choosing a modified path in parameters space, we are able to recover a fidelity $F=0.88$ in the final state, with $F\geq0.85$ at all times. 
This result is a strong evidence that the main ideas used to find the protocol in Fig.~\ref{protocol} are actually quite general and can be applied to other scenarios, i.e. that accessing substantially higher values of total angular momentum allows for faster Laughlin state preparation.

\section{Conclusions} \label{Conclusions}

In this work, we have proposed a time efficient adiabatic protocol to prepare the $\nu=1/2$ fractional quantum Hall ground state of four bosonic atoms. Starting from a condensate in the lowest Landau level, we reach the Laughlin state within $T=605$ trapping periods and with a fidelity of $0.99$.

Our total time of $T=605$ trapping periods represents an improvement by a factor of $10$ when compared to the 6450 trapping periods in Ref.~\cite{popp04}. For a trapping frequency of $(2\pi)\times$30~kHz, our  protocol would take only $20~$ms. However, the experimental work of Ref.~\cite{gemelke10} considers a trapping frequency of only $(2\pi)\times$2~kHz, for which our protocol would take $300~$ms. The former value sets a feasible time scale for the adiabatic preparation of correlated states with cold atom systems, which always has to be balanced with intrinsic heating rates. Consequently, the presented results will be valuable in guiding experiments with cold atoms aiming at the preparation of Laughlin states with rotating mini-traps. 

An important feature of our protocol is usage of large anisotropies \footnote{In this context, we stress the different definition of our parameter $A$ as compared to the anisotropy parameter $\epsilon$ used in Ref.~\cite{popp04}.}.  This leads to ellipticities which in our protocol are twice as large as in Ref.~\cite{popp04}. The correct description in the regime of large deformation is numerically expensive, but our study shows that strong anisotropy is important for reaching fast adiabatic ramps. Large rotating quadrupolar deformations are experimentally feasible be it in optical traps \cite{gemelke10}, in a time-orbiting potential trap \cite{fletcher2019} or by a rotating pair of repulsive optical traps \cite{Abo-Shaeer2001}.

For an accurate description in
the vicinity of the centrifugal limit, we had to ensure a sufficiently
large angular momentum truncation $L_{\rm max}$: The contour plots of the energy gap, Fig.~\ref{colorplots}, considerably depend on this truncation.  In particular, by truncating at the low value of $L_{\rm max}=12$, the Laughlin region (lower right corner of the contour plot) appears fully separated from other regions by a valley of very small energy gap. This hinders the fast preparation of the Laughlin state. Allowing for larger angular momentum changes this picture, and the Laughlin state can then be reached without crossing such a valley of small gaps, if the anisotropy parameter is chosen sufficiently large. 

In this work, we have assumed an interaction parameter of $g=1$. This is slightly larger than the value $g=0.6$ assumed in Ref.~\cite{popp04}, or  $g=0.41$ in Ref.~\cite{gemelke10}. Sufficiently strong interactions are important because the many-body gap above the Laughlin state scales as $\sim0.1g\:\hbar\omega$ \cite{regnault2003,julia-diaz12}. While many experiments operate in the weakly-interacting regime with $g\approx0.1$, strong interactions of $g\approx3$ have been realized using a Feshbach resonance \cite{Ha2013}. In principle, it is also possible to tune $g$ as a function of time. This would provide another knob in the state preparation scheme - an opportunity which is left for future work.

We expect that, if experimentally required, the preparation time can be further reduced. An adiabatic scheme could for instance benefit from exploring even larger anisotropies, or from introducing more points $P_i$ at which ramps are changed. In this context, optimal control strategies for many-body systems \cite{doria11} might be used to find the best path, however, in practice, this possibility is limited by the fact that simulating systems with large ellipticities is numerically expensive. Such optimization protocols might also leave behind adiabatic  paths, and it would be interesting to investigate whether counter-diabatic preparation schemes can achieve better results.

\begin{acknowledgements}
The authors would like to thank Klaus Sengstock, Leticia Tarruell, Fabian Grusdt and Bruno Juli\'{a}-D\'{i}az for fruitful discussions and helpful comments.
BA acknowledges funding from the European Union's Horizon 2020 research and innovation programme under the Marie Sk{\l}odowska-Curie grant agreement No. 847517, Maria Yzuel Fellowship and Coordenação de Aperfeiçoamento de Pessoal de Nível Superior - Brasil (CAPES) - Finance Code 001.
T.G. acknowledges financial support from a fellowship granted by “la Caixa” Foundation (ID 100010434, fellowship code LCF/BQ/PI19/11690013).
B.A., V.K., M.L., and T.G. acknowledge funding from  ERC AdG NOQIA, Spanish Ministry MINECO and State Research Agency AEI (FISICATEAMO and FIDEUA PID2019-106901GB-I00/10.13039 / 501100011033,
SEVERO OCHOA No. SEV-2015-0522 and CEX2019-000910-S, FPI), European Social Fund, Fundaci\'o Cellex, Fundaci\'o Mir-Puig, Generalitat de Catalunya (AGAUR Grant No. 2017 SGR 1341, CERCA program, QuantumCAT U16-011424, co-funded by ERDF Operational Program of Catalonia 2014-2020), MINECO-EU QUANTERA MAQS (funded by State Research Agency (AEI) PCI2019-111828-2/10.13039/501100011033), EU Horizon 2020 FET-OPEN OPTOLogic (Grant No 899794), and the National Science Centre, Poland-
Symfonia Grant No. 2016/20/W/ST4/00314.
C.W. acknowledges funding from the European Research Council (ERC) under the European Union's Horizon 2020 research and innovation programme under grant agreement No. 802701.
\end{acknowledgements}

\appendix
\section{Comparison to Ref.~\cite{popp04}}
\label{AppendixPopp2004}

We dedicate this section to analyse what makes our work different from Ref.~\cite{popp04} by Popp, Paredes, and Cirac. First, we explain how to compare the interaction term used in our work to the one in \cite{popp04}. Then we use their model and same parameters to reproduce their energy gap plot, unfortunately the authors do not mention the exact value used for angular momentum truncation. In any case, we give arguments to convince the reader that the total angular momentum truncation used by the authors of Ref.~\cite{popp04} was not enough to represent the states along their protocol. Finally, we provide an explanation why the lower choice of total angular momentum truncation prevented their protocol to the Laughlin state from being faster. More precisely, their protocol is ten times longer than ours.

The model used by the authors is given by the Hamiltonian
\begin{align}
    H = \left(1-\frac{\Omega}{\omega}\right) L + 2\pi\eta\:U + V'(t).
\end{align}
They use the same interaction potential $U$, but with different form of the coupling constant multiplying it. They work with $\eta=0.1$, which corresponds to $g=0.63$ in our model. The anisotropic potential used in their work is
\begin{align}
    V'(t) \propto \omega^2 (1+\epsilon)^2 x^2 + \omega^2 y^2,
\end{align}
where $\epsilon$ is the small anisotropic parameter. In this case, the anisotropy is due to an increase of the trapping frequency along the $x$-direction, while no change is made in the trapping frequency along the $y$-direction. For this reason, there will be no region that requires more delicate preparation in their energy gap plots, differently from what we had in Fig.~\ref{protocol}. In terms of Fock space operators, and in units of $\hbar\omega$, their potential is
\begin{align}
    V'(t) =& \:\:\frac{\epsilon}{4}\sum_{m=2}^\infty \left[\sqrt{m(m-1)}a^\dagger_m a_{m-2}+{\rm h.c}.\right] \nonumber \\
    &+\frac{\epsilon}{2}\sum_{m=0}^\infty(m+1)a^\dagger_m a_m.
\end{align}
This is the same as our potential, but with an extra diagonal term in the second line. We see that the trap deformation of our work is two times stronger than what they used. 

\begin{figure}[t] 
  \includegraphics[scale=0.9]{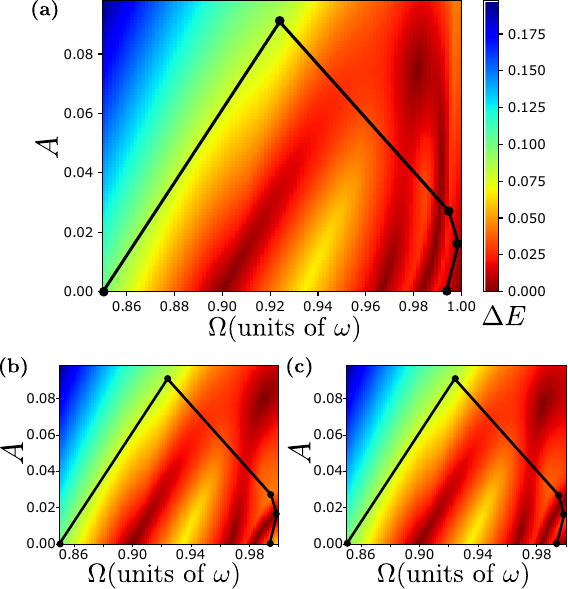}
  \caption{Energy gap plots obtained by using the same Hamiltonian as in \cite{popp04} with different values of angular momentum truncation. (a) $L_{\rm{max}}=40$, (b) $L_{\rm{max}}=12$, and (c) $L_{\rm{max}}=14$. Comparing with the energy gap plot in \cite{popp04}, we believe they used either $L_{\rm{max}}=12$ or $14$. The black curve is an approximation of the path proposed in Ref.~\cite{popp04}, time dependence and exact coordinates of the points were not provided by the authors. An improved path should avoid the narrow gap region by accessing large anisotropy, $A>0.09$, and then decrease it only after reaching around $\Omega=0.99$.}
  \label{PoppPlots}
\end{figure}

In Fig.~\ref{PoppPlots}, we show our attempts to reproduce their energy gap plots. Since the authors do not give the value used for angular momentum truncation, we ran simulations for different values of $L_{\rm{max}}$, and based on the images we believe that they used either $L_{\rm{max}}=12$ or $L_{\rm{max}}=14$. When used $L_{\rm{max}}=40$, however, the energy gap plot becomes very different, and it becomes clear that there is actually a path that leads to the Laughlin state without need to cross the narrow gap region. Using $L_{\rm{max}}=12$ or $L_{\rm{max}}=14$, one is induced to think that the only possibility to reach the Laughlin state is by crossing the narrow gap region, which caused their protocol to be extremely long. The protocol in Ref.~\cite{popp04} cannot be considered realistic as they are missing an important part of the Hilbert space by truncating the total angular momentum in such low values.

In Ref.~\cite{popp04}, the authors propose protocols for preparing ground states with angular momentum $L=4$ and $L=8$. The suggested protocol prepares the $L=4$ ground state in $240$ trapping periods, and the $L=8$ in $360$ periods, both with fidelity $F=0.99$. The authors of Ref.~\cite{popp04} indicate the geometrical shape of the path, but not its parametrization. We also obtained protocols with high fidelities for the intermediate ground states of our model, but without a significantly improved preparation times in comparison to Ref.~\cite{popp04}. The $L=4$ ground state was prepared in $160$ trapping periods with final fidelity $F=0.99$, while the $L=8$ state was prepared in $320$ trapping periods with final fidelity $F=0.95$.

\bibliography{bibliography.bib}

\end{document}